\documentclass[journal]{IEEEtran}
\usepackage{amsmath}
\usepackage{amsfonts}
\usepackage{graphicx}
\usepackage{balance}
\usepackage{xcolor}
\usepackage{glossaries}

\usepackage{siunitx}
\setacronymstyle{long-short}
\newacronym{pl}{PL}{photonic lantern}
\newacronym{mspl}{MSPL}{mode-selective photonic lantern}
\newacronym{itr}{ITR}{inverse taper ratio}
\newacronym{lp}{LP}{linearly polarized}
\newacronym{dcf}{DCF}{double-clad fiber}

\newcommand\unmarkedfootnote[1]{
    \begingroup
        \renewcommand\thefootnote{}\footnote{#1}
        \addtocounter{footnote}{-1}
    \endgroup
}

\begin{document}

\title{Mode-Selective Photonic Lanterns\\with Double-Clad Fibers}

\author{
\IEEEauthorblockN{
    Rodrigo Itzamn\'a Becerra-Deana\IEEEauthorrefmark{1}\IEEEauthorrefmark{2},
    Martin Poinsinet de Sivry-Houle\IEEEauthorrefmark{1},\\
    St\'ephane Virally\IEEEauthorrefmark{1}\IEEEmembership{Member, IEEE},
    Caroline Boudoux\IEEEauthorrefmark{1}\IEEEauthorrefmark{2}
    and Nicolas Godbout \IEEEauthorrefmark{1}\IEEEauthorrefmark{2}\IEEEauthorrefmark{3}
}\\
\IEEEauthorblockA{
    \IEEEauthorrefmark{1}Polytechnique Montr\'eal, 2500 Chemin de Polytechnique, Montr\'eal, QC H3T 1J4, Canada}\\
\IEEEauthorblockA{
    \IEEEauthorrefmark{2}Castor Optics, 361 Boulevard Montpellier, Saint-Laurent, QC H4N 2G6, Canada}\\
\IEEEauthorblockA{
    \IEEEauthorrefmark{3}nicolas.godbout@polymtl.ca}
}


\maketitle
\unmarkedfootnote{Caroline Boudoux and St{\'e}phane Virally acknowledge funding from the Mid-Infrared Quantum Technology for Sensing (MIRAQLS) project, supported by the European Union’s Horizon Europe research and innovation programme under grant agreement 101070700.}

\begin{abstract}
    We present the design, fabrication, and characterization of mode-selective photonic lanterns using double-clad fibers. Here, we exploited several custom-pulled double-clad fibers to achieve the symmetry break required to excite higher-order modes. 
    The resulting components are short and exhibit high modal isolation and low excess loss.
    They address some of the limitations of existing photonic lanterns in terms of fragility and coupling efficiency.
    The fabrication process involves the use of lower-index capillary tubes to maintain fiber geometry during fusion and tapering.
    Through the use of varying first cladding diameters, mode selectivity is achieved without sacrificing single-mode compatibility.
    This in turn allows proper real-time characterization during the whole fabrication process.
    Results demonstrate that double-clad fibers stacked inside a fluoride-doped capillary tube feature high modal isolation (above 60~\unit{dB}) and low excess loss (lower than 0.49~\unit{dB}), over a broad wavelength range (more than 250~\unit{nm}) with steeper taper profiles, and more robust components.
    The use of less expensive synthetic fused silica capillary tubes achieves high modal isolation (above 20~\unit{dB}) and excess loss lower than 2~\unit{dB} over the same broad wavelength range. 
    
\end{abstract}


\section{Introduction}\label{sec:Introduction}

    Fiber-based space-division multiplexers, also known as \glspl{pl}, were introduced in 2005 as reversible multimode to single-mode multiplexers~\cite{leon-saval_multimode_2005}.
    These fiber-based devices are designed for streamlined integration into compact, alignment-free optical systems.
    They are widely used across multiple disciplines, including astrophysics, telecommunications, and biophotonics, and for many applications such as illumination, spectroscopy, imaging, and efficient scattering collection~\cite{olaya_161_2012, leon-saval_photonic_2013, betters_beating_2013, ozdur_free-space_2013, velazquez-benitez_optical_2018, sivry-houle_all-fiber_2021, Maltais-Tariant:23, Raphael22}.
    They are usually fabricated by fusing and tapering several ($N\geq2$) fibers to generate a few-mode or multimode guiding structure with a few tens of micrometers diameter.
    The plurality of design possibilities gives rise to distinct types of \glspl{pl}.
    Among them, \glspl{mspl} act as mode sorters, allowing the guided transverse modes of the multimode section to be selectively coupled to a specific single-mode fiber~\cite{shen_highly_2018}.
    \glspl{mspl} usually feature high modal isolation, and their operation is broadband, up to a bandwidth of 500~\unit{nm}~\cite{fontaine_photonic_2022}, with excess loss usually lower than 1~\unit{dB}~\cite{birks_photonic_2015}.
    
    Some applications of \glspl{pl}, such as biomedical imaging and sensing, require sturdy devices capable of being embedded in portable packages.
    This, in turn, implies the design of shorter components, as long fiber components tend to be much more fragile.
    The limiting factor in such designs is the adiabatic criterion~\cite{deSivry-Houle:24}: a number characterizing the steepest slope at any taper point that allows an adiabatic transfer of modes. 
    A higher adiabatic criterion allows steeper slopes, leading to shorter components.
    Previous attempts at reducing the length of \glspl{pl} have included designs at the limit of existing adiabatic criterion graphs with regular single-mode fibers~\cite{sunder_engineering_2021}.
    Unfortunately, this technique is limited by current single-mode fiber technology. 
    Another way is to use graded-index fibers~\cite{huang_all-fiber_2015-2} with logarithmic refractive index distributions~\cite{harrington_endlessly_2017}. 
    This has proven effective in increasing the adiabatic criterion and thus reducing the size of the devices.
    
    \glspl{mspl} require symmetry breaks to enable one-to-one mapping of the transverse modes in the few-mode section to specific single-mode fibers.
    Thus, an $N\times1$ \gls{mspl} will require $N$ different fibers.
    This is difficult to achieve with graded-index fibers without compromising compatibility with the single-mode fibers used everywhere in full imaging or sensing systems~\cite{adams_splicing_2008}.
    Hence, \glspl{mspl} made of currently available graded-index fibers remain limited to few-mode structures.

    In this paper, we demonstrate the use of \glspl{dcf} with single-mode cores to design and fabricate \glspl{mspl} with increased adiabatic criteria and excellent compatibility with single-mode fibers.
    The components fabricated using this technique prove to be short and sturdy.
    They also feature high modal isolation and low excess loss.

\section{Design and fabrication}\label{sec:Methods}

\begin{figure}[!t]
        \centering
        \includegraphics[width=3in]{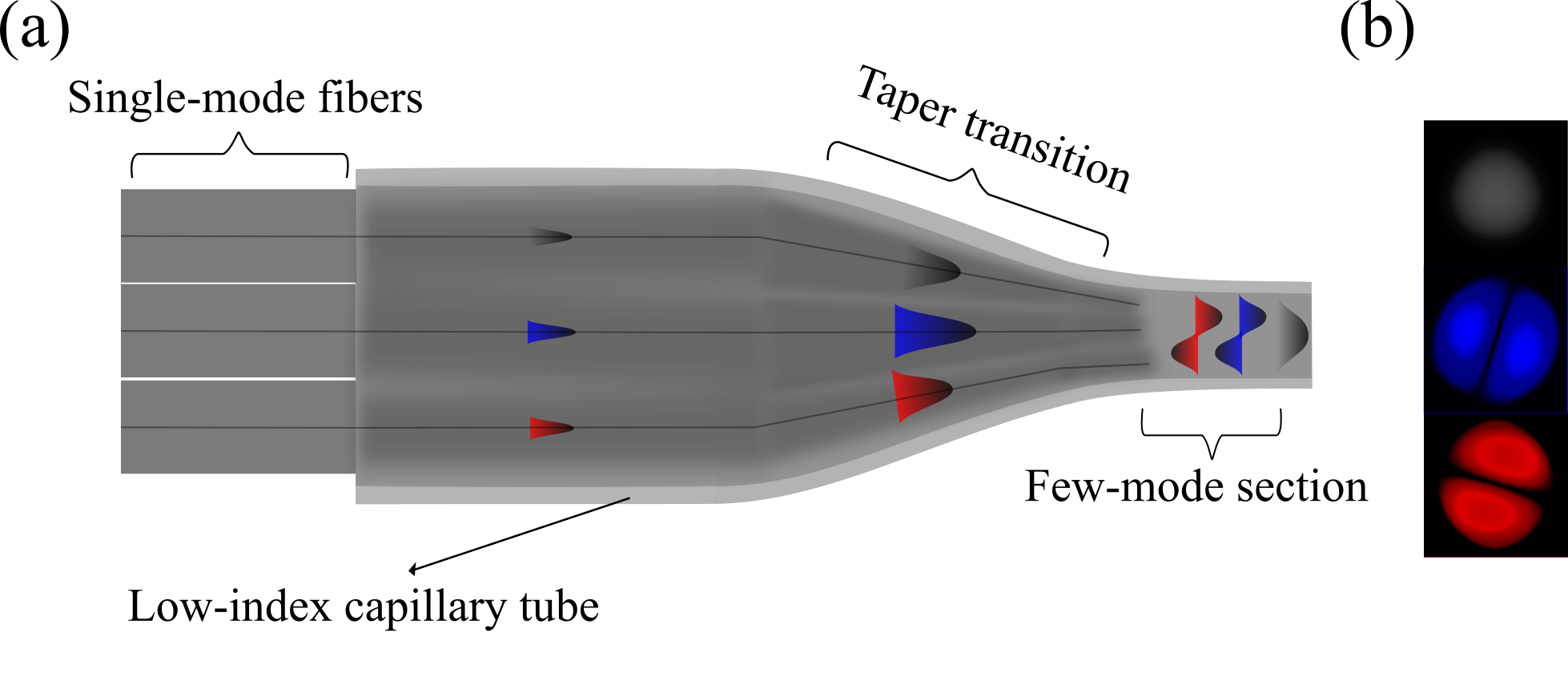}
        \caption{Schematic of a $3\times1$ photonic lantern.
        (a) From left to right: three single-mode fibers are inserted into a low refractive index capillary tube before being fused and tapered down towards the few-mode section.
        Image not to scale: the outer diameter of the single-mode fibers is 125~\unit{\micro\meter}, while the outer diameter of the few-mode structure is approximately 10~\unit{\micro\meter}. (b) Optical profile of the three output modes akin to LP$_{10}$ (black), LP$_{11\mathrm{a}}$ (blue), and LP$_{11\mathrm{b}}$ (red).}
        \label{fig:lantern_schematics}
    \end{figure}

    \Glspl{mspl} are composed of three main sections: a single-mode fiber bundle, a tapered transition, and a few- or multimode structure.
    Figure~\ref{fig:lantern_schematics} presents the longitudinal view of a $3\times1$ \gls{mspl} encased inside a fluoride-doped capillary tube to constrain the fiber geometry during the fusion and tapering process.

\subsection{Design}\label{sec:design}

    The design of \glspl{pl} involves a large number of degrees of freedom, including the number of fibers, their types, the stacking configuration, the degree of initial fusion, and the tapering profile~\cite{fontaine_photonic_2022, fontaine_geometric_2012, tedder_single-mode_2019, davenport_optimal_2021}. 
    In an \gls{mspl}, each fiber in the single-mode bundle must exhibit a distinct refractive index profile to break the symmetry and allow to selectively excite individual transverse modes within the few-mode structure.
    Given the size of the design parameter space, we use SuPyMode~\cite{deSivry-Houle:24}, a Python-based simulation tool that uses coupled-mode theory to efficiently compute the adiabatic criterion given the characteristics of the input fiber bundle and the initial degree of fusion. SuPyMode allows rapid iteration and convergence towards adequate fiber parameters for most applications.
    
    \Glspl{dcf} can be designed to offer full compatibility with single-mode fibers.
    They offer an important advantage over single-mode fibers used in some applications: when \glspl{dcf} are tapered, the lower-order modes remain contained inside the first cladding even as the core ceases to guide them.
    In applications such as \glspl{pl}, this containment mechanism helps to increase the adiabatic criterion~\cite{deSivry-Houle:24, guo_ultra-low-loss_2023, Rodrigo_SPIE, Zhang:24}, albeit to a lesser degree than with graded-index fibers.
    Containment in \glspl{dcf} can be achieved with a large variety of first-cladding diameters without changing the diameter of the core.
    This greatly increases the versatility of designs, while preserving full compatibility with single-mode fibers, a feature difficult to replicate with graded-index fibers.
        
    \begin{figure*}[!t]
        \centering
        \includegraphics[width=\textwidth]{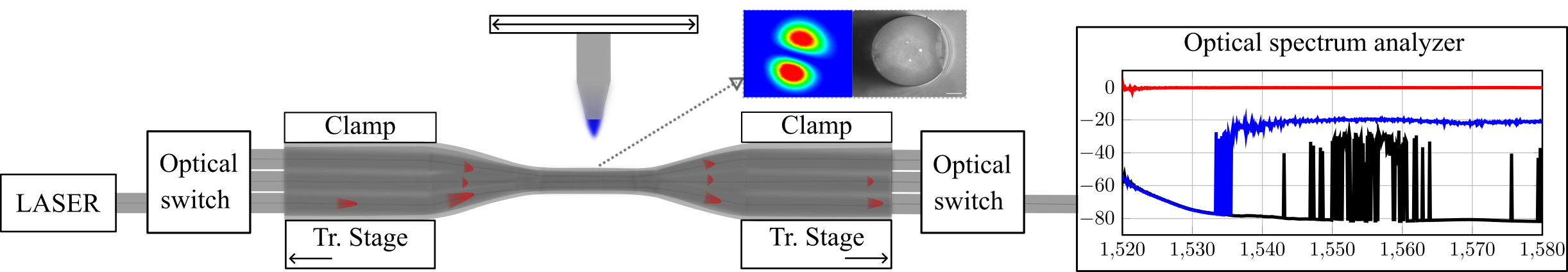}
        \caption{Schematic diagram of the fusion tapering station, comprising a flame, clamps, and translation stages, attached to a full \textit{in situ} characterization system comprising a laser, optical switches, and an optical spectrum analyzer. Optical switches allow for all inputs to be illuminated and all outputs to be monitored independently. Inset: one of the LP$_{11}$ modes (left) and image of the few-mode section, both observed post cleave.}
        \label{fig2}
    \end{figure*}

\subsection{Fabrication}\label{sec:fab}
    Low-index capillary tubes are used during the fabrication process to constrain the fibers' geometry and provide additional containment in the tapered section.
    A capillary tube is first tapered in its center using an automated glass processing system (GPX-3000 Vytran, Thorlabs, NJ, USA) with an FTAT4 filament (Thorlabs, NJ, USA).
    In a $3\times1$ \gls{mspl}, the tube is tapered down to an inner diameter of 270~\unit{\micro m} to constrain the three-fiber bundle into an equilateral triangle configuration over a length of 20~\unit{mm} to provide adequate scanning length.
    Once the proper fiber configuration is obtained, a fusion step is performed, which slightly decreases the bundle size.
    The tapering process further reduces the few-mode or multimode part of the \gls{pl} to its final size, around 10 to 20~\unit{\micro m}. The single-mode compatibility of the \glspl{dcf} fibers with SMF28 allows monitoring the fabrication process in real-time by injecting light into any input and measuring the power at all outputs.
    
    Figure~\ref{fig2} shows the in-situ characterization process of a $3\times1$ \gls{mspl}, with inputs connected through an in-house built optical switch to a broadband laser (BBS1550, JDS Uniphase, AZ, USA), centered at 1550~\unit{nm} with a bandwidth of 50~\unit{nm}.
    The component is placed on a heat-and-pull fabrication station, which consists of two holders with homemade clamps mounted on translation stages symmetrically arranged around a moving torch.
    Outputs are connected to an optical switch (built in-house) to monitor output sequentially on an optical spectrum analyzer (OSA, model AQ6317, Ando, Japan).
    This setup allows modal isolation and excess loss to be monitored in real-time after establishing a baseline.
    Any loss due to micro-deformations, dust, and unwanted coupling between modes can be observed in real-time, and the recipe can be altered accordingly, or the process can be halted immediately.

\section{Results}\label{sec:Results}

    Here, we describe the design, fabrication, and characterization of $3\times1$ \glspl{mspl} using \glspl{dcf}.
    Three distinct \glspl{dcf} were custom-designed to offer single-mode operation at 1550~\unit{nm}, matching the mode-field diameter of telecommunication (SMF-28) fibers.
    The fibers featured three different first cladding diameters (19.6, 26.8, and 42~\unit{\micro\meter}, respectively) without noticeable differences in their single-mode operation and with a numerical aperture of 0.13 between the core and first cladding.
    Slight differences in doping concentrations allowed the multimode guiding part (i.e., within the first to second cladding) to feature the same numerical aperture (0.117) for all three fibers. 

    \begin{figure}[!ht]
        \centering
        \includegraphics[width=\columnwidth]{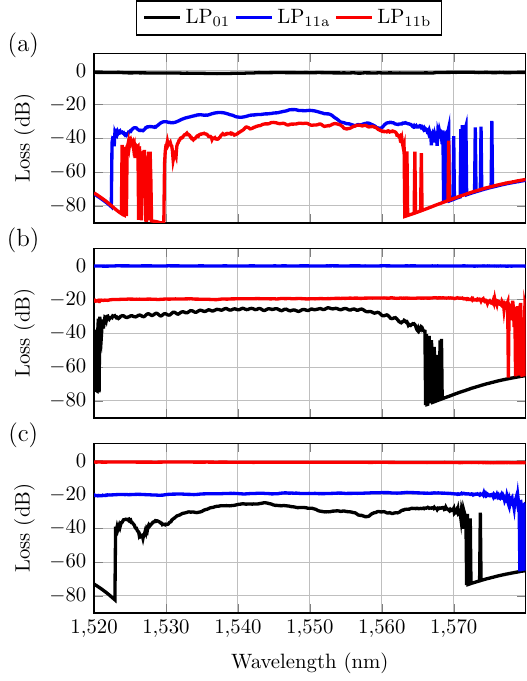}
        \caption{Differential power measurements at the output of a $3\times1$ lantern inside a capillary tube made of synthetic fused silica with a slightly lower index of refraction than that of the outer cladding of the fibers.
        The measurement is made at the end of the fusion-taper process, before cleaving.
        (a): core illumination through the fiber with the largest first-cladding diameter, resulting in the fundamental mode, akin to LP$_{01}$, in the few-mode section.
        (b): core illumination through the fiber with the second largest first-cladding diameter, which generates the next mode, akin to LP$_{11\mathrm{a}}$.
        (c): illumination through the fiber with the smallest first-cladding diameter, which generates the last guided mode, akin to LP$_{11\mathrm{b}}$.}
        \label{fig:osa_measurements_vitron}
    \end{figure}  

    \begin{figure}[!ht]
        \centering
        \includegraphics[width=\columnwidth]{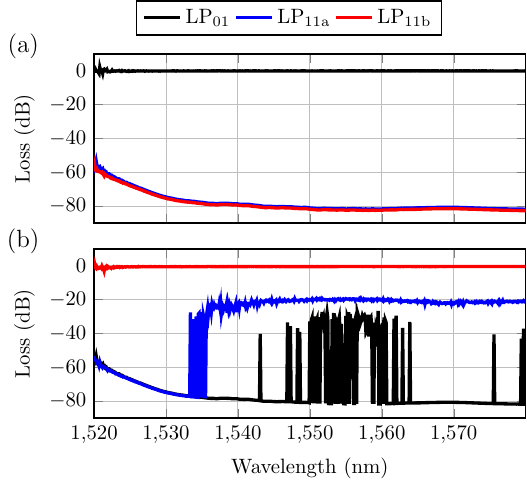}
        \caption{Differential power measurements at the output of a $3\times1$ lantern inside a capillary tube made of fluoride-doped silica, with an even lower index of refraction than that of the outer cladding of the fibers.
        (a): core illumination through the fiber with the largest first-cladding diameter, resulting in the fundamental mode, akin to LP$_{01}$, in the few-mode section.
        (cb: illumination through the fiber with the smallest first-cladding diameter, which generates the last guided mode, akin to LP$_{11\mathrm{b}}$.
        An accidental break of the third fiber resulted in it not being illuminated during that run.}
        \label{fig:osa_measurements}
    \end{figure}
    
    Figure~\ref{fig:osa_measurements_vitron} shows the optical power at each output of the $3\times1$ MSPL for illumination in each input, at the end of the tapering process.
    The MSPL characterized in Fig.~\ref{fig:osa_measurements_vitron} was realized with the DCFs described above, and using a synthetic fused silica capillary tube (CV1012, Vitrocom, NJ, USA). Fig.~\ref{fig:osa_measurements_vitron}(a) shows the power transfer for illumination into Input 1 consisting of the DCF with the largest inner cladding, expected to excite a supermode akin to the first linearly-polarized (LP) mode, LP$_{01}$, in the fused section.
    Results show power transfer with $\le1.84~\unit{dB}$ excess loss into Output 1 (black curve), and $\ge20~\unit{dB}$ extinction in Output 2 (blue curve) and Output 3 (red curve) ports, respectively. 
    Similarly, Fig.~\ref{fig:osa_measurements_vitron}(b) shows the power transfer for illumination into Input 2, the DCF with the second largest inner cladding, expected to excite a supermode akin to LP$_{11a}$ in the fused section.
    Results show quasi-lossless transfer ($\le0.64~\unit{dB}$ excess loss) into Output 2 (blue curve), and $\ge20~\unit{dB}$ extinction in Outputs 1 (black curve) and 3 (red curve), respectively. 
    Finally, Fig.~\ref{fig:osa_measurements_vitron}(c) shows the power transfer for illumination into Input 2, the DCF with the smallest inner cladding, expected to excite a supermode akin to LP$_{11b}$ in the fused section.
    Results show power transfer with $\le1.65~\unit{dB}$ excess loss into Output 3 (red curve), and $\ge20~\unit{dB}$ extinction in Outputs 1 (black curve) and 3 (red curve), respectively.

    Figure~\ref{fig:osa_measurements} shows the optical power at each output of the $3\times1$ MSPL for illumination in two different inputs, right at the end of the tapering step.
    For this MSPL, the three DCFs were arranged inside a fluoride-doped fused silica capillary tube (FTB03, Thorlabs, NJ, USA). Fig.~\ref{fig:osa_measurements}(a) shows the power transfer for illumination into Input 1, the DCF with the largest inner cladding, expected to excite a supermode akin to LP$_{01}$ in the fused section.
    Results show power transfer with $\le0.49~\unit{dB}$ excess loss into Output 1 (black curve), and $\ge60~\unit{dB}$ extinction in Output 2 (blue curve) and Output 3 (red curve) ports, respectively. 
    Similarly, Fig.~\ref{fig:osa_measurements}(b) shows the power transfer for illumination into Input 3, the DCF with the smallest inner cladding, expected to excite a supermode akin to LP$_{11b}$ in the fused section.
    Results show quasi-lossless transfer ($\le0.46~\unit{dB}$ excess loss) into Output 2 (blue curve), and $\ge20~\unit{dB}$ extinction in Outputs 1 (black curve) and 3 (red curve), respectively. 
    Power transfer for illumination into Input 2 was not recorded for this component.
    However, the coupling between the two branches is the same in both directions: illuminating Input X and measuring the loss in Output Y gives the same information as illuminating Input Y and measuring the loss in Output X.
    This can indeed be observed in Fig.~\ref{fig:osa_measurements_vitron}, e.g. comparing the red trace in Fig.~\ref{fig:osa_measurements_vitron}(b) to the blue trace in Fig.~\ref{fig:osa_measurements_vitron}(c).
    In that respect, apparently anomalous points in the black curve of Fig.~\ref{fig:osa_measurements}(b), which should follow the red curve of Fig.~\ref{fig:osa_measurements}(a), and does so most of the time, are probably due to micro stresses (fibers moving) during the fast measurement.
    In the end, the only information missing from the lack of illumination in Input 2 is the excess loss for the corresponding fiber.
    It is, however, expected to be in line ($\le0.5~\unit{dB}$) with that of the fibers corresponding to Inputs 1 and 3, as is usually the case.
    
    \begin{figure}[!t]
        \centering
        \includegraphics[width=\columnwidth]{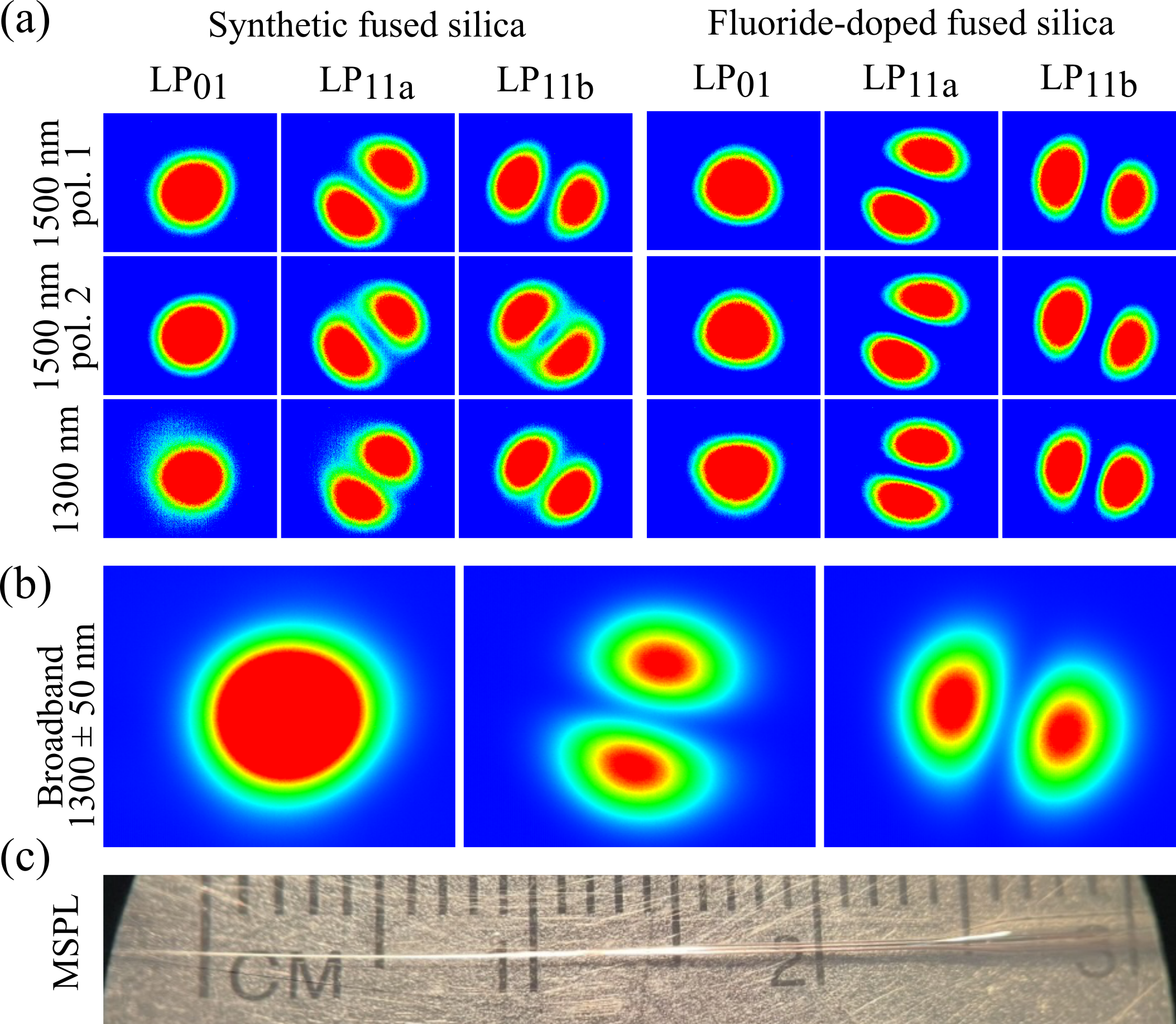}
        \caption{Far-field images of the modes at the output of the few-mode section, illuminating through each of the single-mode fiber input to generate a) LP${01}$, b) LP$_{11a}$, and c) LP$_{11b}$. d) Picture of the photonic lantern under.}
        \label{fig4}
    \end{figure}

    After the fusion and tapering process, both structures were cleaved, and each input port was individually illuminated to capture an optical profile.
    A fiber polarization controller was used to characterize the component's behavior with respect to input polarization. 
    An image in the far field was obtained by projecting each output onto a screen and recording using an infrared camera (SU320KTS-1.7RT, Goodrich, NJ, USA).
    Narrow-band illumination at 1300~\unit{nm} and 1550~\unit{nm} was achieved with a dual-wavelength source (HP 8153ASM, Hewlett-Packard, CA, USA). A broad source (100~\unit{nm} wide around 1300~\unit{nm}) was used for broadband illumination (SL1310V1-10048, Thorlabs, NJ, USA).
    
    Figure~\ref{fig4}(a) shows the optical profile in the far field of an MSPL based on a synthetic fused silica capillary (left) and based on a fluoride-doped fused silica capillary (right), respectively. For each group, the left column shows the fundamental mode of the structure (akin to LP$_{01}$) when the largest first-cladding diameter is illuminated. The center and right columns, respectively, show the two other modes (akin to LP$_{11a}$ and LP$_{11b}$, respectively) in orthogonal orientations when the other two inputs are illuminated.
    The first two rows correspond to narrow-band illuminations at 1550~\unit{nm}, with two different polarization states finding the most significant difference between each mode.
    They show that the behavior of the components is independent of the polarization state of the illumination source.
    The third row corresponds to narrow-band illumination at 1300~\unit{nm}.
    The component essentially behaves independently of the illumination wavelength. 
    This conclusion is reinforced by the result of Figure~\ref{fig4}(b), which shows the optical profile using broadband illumination around 1300~\unit{nm} for LP$_{01}$, LP$_{11a}$, and LP$_{11b}$, respectively, for the fluoride-doped capillary tube MSPL.
    Figure~\ref{fig4} (c) shows the tapered length of an MSPL after cleaving in the middle of the tapered section.
    It shows that the full transition between the single-mode bundle and the few-mode section is around 2.5~\unit{cm}.

\section{Discussion}

    The components fabricated for this study exhibited excess low loss: $<1~\unit{dB}$ for the components held in a fluoride-doped capillary tube and $<2~\unit{dB}$ for the component held in synthetic fused silica.
    They also feature large modal isolation above $20~\unit{dB}$, and in the case of the fluoride-doped components above $60~\unit{dB}$, the measurement being limited only by the noise floor of the measuring device.
    
    Proper symmetry breaking was obtained with differences of first-cladding diameter as low as 6~\unit{\micro m}.
    From simulations, we believe that differences down to 4~\unit{\micro m} would be sufficient, but experimental confirmation is required.

    For both components characterized in Figures~\ref{fig:osa_measurements_vitron} and ~\ref{fig:osa_measurements}, the operational bandwidth was limited by the bandwidth of the illumination laser (JDS Uniphase 1550~\unit{nm}).
    As such components do not rely on mode coupling, they are theoretically as broadband as the guiding properties of the fibers allow.
    A comparison of the results between the two components shows that using a higher index difference between the outer cladding of the fibers and the external capillary tube usually leads to better modal isolation and lower excess losses.
    Although the fluoride-doped capillary tubes give overall better results, they are much more expensive than those made of synthetic fused silica.
    
    In principle, the fabrication process enables two lanterns to be extracted from each fused and tapered bundle, by cleaving exactly in the middle of the tapered section.
    However, the component is not fully symmetric to start with. Fiber insertion within the capillary tube requires stripping the coating over a distance of at least 20~\unit{cm} longer than the capillary, on one of its sides. For two lanterns to be made out of the component, re-coating of that section is required to avoid fragility.

    In theory, the fabrication of \glspl{mspl} should never result in fiber-to-fiber coupling, in contrast to what is expected for couplers.
    Any such power transfer is thus indicative of a flaw in the device or a non-adiabatic section caused by too steep a transition (e.g., a micro bend) in the device.
    Real-time monitoring enables these types of flaws to be immediately detected.
    In-situ monitoring also enables the measurement of actual mode isolation and excess loss, contrary to the cutback method, which measures losses from each input into the few-mode section after cleaving.
    This method widely underestimates excess losses as it does not measure cladding or coupling losses.
    For instance, for the component made with a synthetic silica tube, we measured excess losses of 1.84, 0.64, and 1.65~\unit{dB} for Inputs 1,2 and 3, respectively.
    But the cutback technique, performed after cleaving, gave 0.08, 0.10, and 0.11~\unit{dB} for the same Inputs, respectively. 

    Double-clad fiber-based MSPLs allow for high modal isolation; however, the resulting modes are not quite perfectly matched to the theoretical LP modes. Some mode asymmetry was observed, likely due to the final shape of the few-mode section, which is not perfectly cylindrical.

\section{Conclusion}\label{sec:Conclusion}
    The use of \glspl{dcf} in the design of \glspl{mspl} provides a versatile means to break symmetries by varying the first-cladding diameter while retaining full compatibility with single-mode fibers.
    This allows for complete characterization of the devices in real-time during fabrication.
    
    The adiabatic criteria of the \gls{dcf}-based components is slightly lower than those made of graded-index fibers.
    However, they are high enough that short ($\simeq2.5~\unit{cm}$), sturdy components can be fabricated with relative ease.
    Our experiments confirm that \glspl{mspl} in general keep their characteristics over large bandwidths (at least 250~nm in this study) and are agnostic to the state of polarization of the illumination.

    Components held in synthetic fused silica capillary tube have higher overall excess losses and lower modal isolation figures than those held in fluoride-doped capillary tubes.
    However, the former are much cheaper than the latter, and they provide a good alternative for components that do not require the best specifications. 

\section*{Disclosures}
RIBD: Castor Optics, inc. employee, CB: Castor Optics, inc. co-presidents, co-founders, and co-owners, NG: Castor Optics, inc. co-presidents, co-founders, co-owners, and patent

\bibliographystyle{IEEEtran}
\bibliography{bibliography}

\begin{IEEEbiographynophoto}{Rodrigo Itzamn\'a Becerra-Deana}
    is pursuing a Ph.D. in the Fiber Optics Laboratory and the Laboratory for Optical Diagnoses and Imaging in the Engineering Physics department at Polytechnique Montr\'eal. He is also part of Castor Optics, where he develops novel fiber components.
\end{IEEEbiographynophoto}\vskip -2.5\baselineskip plus -1fil
\begin{IEEEbiographynophoto}{Martin Poinsinet de Sivry-Houle}
    is Research Fellow in the Engineering Physics department at Polytechnique Montr\'eal. He developed SuPyMode, the library used in all the simulations for the design of fiber components in the Fiber Optics Laboratory.
\end{IEEEbiographynophoto}\vskip -2.5\baselineskip plus -1fil
\begin{IEEEbiographynophoto}{St\'ephane Virally}
    is Research Fellow in the Engineering Physics department at Polytechnique Montr\'eal. He is named as co-inventor on over fifteen patents worldwide. He currently leads the Fiber Optics Laboratory.
\end{IEEEbiographynophoto}\vskip -2.5\baselineskip plus -1fil
\begin{IEEEbiographynophoto}{Caroline Boudoux}
    is Professor of Engineering Physics at Polytechnique Montr\'eal, SPIE Fellow, and Director-at-Large at Optica. She founded the Laboratory for Optical Diagnoses and Imaging and co-founded Castor Optics, inc. She is the author of numerous publications, patents, and books on topics ranging from biomedical optics to engineering design and doctoral studies.
\end{IEEEbiographynophoto}\vskip -2.5\baselineskip plus -1fil
\begin{IEEEbiographynophoto}{Nicolas Godbout}
    is Professor and Director of the Engineering Physics department at Polytechnique Montréal. He has led the Fiber Optics Laboratory for more than 15 years and has more than 25 years of experience in the design and fabrication of fiber components for applications in telecommunications, biophotonics and quantum optics. He is named as co-inventor on over ten patent families that are actively exploited. He is also a Co-Founder of Castor Optics which develops and manufactures novel optical fiber components for various markets.
\end{IEEEbiographynophoto}

\end{document}